\documentclass[a4paper,11pt]{article}
\usepackage[english]{babel}
\usepackage[latin3]{inputenc}
\usepackage[T1]{fontenc}
\usepackage{textcomp}
\usepackage{epsfig}
\usepackage{latexsym}
\usepackage{graphicx}
\usepackage{amsfonts,amssymb,bm}
\usepackage{color}

\usepackage[left,pagewise,mathlines]{lineno}
\newcommand{\D}{{\rm d}}

\begin{document}
\title{On the Bell's spaceships paradox and\\
proper length for accelerated bodies}
\author{Jesús Ceresuela and Josep Llosa\\
{\small Departament de Física Quàntica i Astrofísica, Institut de Ciències del Cosmos} \\ Universitat de Barcelona}

\maketitle

\begin{abstract}
We study the Dewan-Beran-Bell thought experiment of two spaceships connected by a thread that start accelerated motion and discuss the proper length of the thread by means of Born's definition of proper length for arbitrary motion. 

\end{abstract}

\section{Introduction}
Although it is commonly known as Bell's spaceships paradox, it was first proposed by Dewan and Beran \cite{Dewan1959} as a puzzle in special relativity aiming to stress the ``physical reality'' of Lorentz contraction. 

Consider a thread hanging between two identical spaceships, $R$ and $F$, that are initialy at rest in some inertial frame $\mathcal{S}\,$, one behind the other a certain distance apart. Simultaneously (as seen from $\mathcal{S}$) they turn on their engines and start moving along the direction $RF$, so acquiring the same accelerations. At any time (in $\mathcal{S}$) their speeds are equal and the distance between $R$ and $F$ as measured by $\mathcal{S}$ keeps constant. According to special relativity, as the overall assembly is moving, the distance measured by $\mathcal{S}$ is affected by Lorentz contraction; thus the thread's [proper] length must increase with speed, so to compensate the Lorentz factor, until the stress associated to that strain is too large and the thread breaks apart. Dewan and Beran \cite{Dewan1959} conclude from this the ``real existence'' of Lorentz contraction, since it ``causes measurable stresses''.

As reported by Bell \cite{Bell}: ``This old problem came up for discussion in the CERN canteen \ldots [and it seemed to emerge] the clear consensus that the thread would not break'' and continues `` \ldots  many people \ldots  get the right answer on further reflection.''

A rather exhaustive sample of the literature on the subject is quoted in ref. \cite{Redzic2008} where Redzic concludes that ``the accepted solution to the riddle (the thread will break when the ships reach a sufficiently high speed), is generally wrong \ldots , while the accepted solution is correct for the mild variant of the problem [at some point the ships run out of gas and continue at a constant speed], it is generally wrong for its tough variant [acceleration never stops]. \ldots according to special relativity in some cases the thread will never break, regardless of how close the ships' speed approaches c.''

This result is rather shocking and brings up some typical traits of paradoxes as it seems to invoke notions which are pushed beyond their limits and that allow some ambiguity because they are not  accurate enough. Think for instance of Lorentz contraction; it establishes the relation between the lengths $L^\dagger$ and $L$ of a stick as measured by two inertial frames, respectively $\mathcal{S}^\dagger$ that sees the stick at rest and $\mathcal{S}$ that sees it in a longitudinal uniform motion at a speed $v$. Both results are different, $\,L=L^\dagger \sqrt{1-v^2/c^2} < L^\dagger\,$, which is natural because both measures imply different procedures \cite{Reichenbach}; while the $\mathcal{S}^\dagger$ measurement only needs geometric  operations, $\mathcal{S}$ has to draw on simultaneity to get a rest ``picture'' of the stick and then apply the above geometric operations on that picture. Besides, if we keep in the framework where it was originally advanced, Lorentz contraction is a sound notion only if uniform rectilinear motions are implied. 

When accelerated motions are involved, things become less simple and one must resort to generalizations, but it may happen that seemingly innocuous assumptions lead to inconsistency.  

We shall distinguish the notions of {\em proper distance} between two events and {\em proper length} of an object. The first of them is the distance between the places where the events happen as measured in an inertial frame in which the events are simultaneous; it only involves two points in spacetime and such a frame always exist, provided that the invariant interval is spacelike. To define the length of a body, e. g. the thread in the Dewan-Beran puzzle, one should resort to an inertial frame in which the whole body is instantaneously at rest, the {\em instantaneous comoving inertial frame} $\mathcal{S}^\dagger$, and then take the length measured in this frame as the body's {\em proper length}.

Very often the latter assumption is too restrictive and, given the motion of a body, an instantaneously comoving inertial frame, that sees the whole body at rest, does not always exist. As proved elsewhere \cite{Llosa2018}, the motions fulfilling this requirement belong to the class of Fermi-Walker motions, and these exhibit the peculiarity that distinct points in the body, that are aligned along the direction of motion, have different proper accelerations. This is not the case for the thread hanging between the spaceships since at least two of these points, $R$ and $F$, have the same acceleration. 

One way to circumvent this obstruction is to take the thread as consisting of infinitesimal pieces; the local instantaneous comoving inertial frame $\mathcal{S}^\dagger$ for each piece is then a frame in which this small part of the thread is at rest in the very instant considered. The {\em proper length} of the piece is defined as the length $\D L^\dagger$ measured in this inertial frame, and it might depend on the instant of time when the measurement is done. Born's definition of length \cite{Born1909} fits in these requirements. Then, to define the proper length of the whole thread we should add (integrate) the infinitesimal proper lengths of all its parts including that, as each $\D L^\dagger$ may be variable over time, this integral should be restricted to {\em simultaneous} infinitesimal measurements. 

This definition involving local instantaneous comoving inertial frames, $\mathcal{S}^\dagger$, it will depend on the motions of the different parts of the thread and, although we shall refer to the {\em proper length of the thread} for shortness, it is rather the proper length of a particular motion of the thread. This definition relies on the so called {\em Hypothesis of Locality} \cite{Mashhoon}  
\begin{quote}
``An accelerated observer measures the same physical results as a standard inertial observer that has the same position and velocity at the time of measurement.''
\end{quote}
One instance of this is the {\em clock postulate} by which a standard clock in arbitrary motion measures {\em proper time}, which is equal to the sum of the infinitesimal amounts of proper time measured by a set of standard comoving inertial clocks: $\; \Delta \tau \int_{t_1}^{t_2} \D t\,\sqrt{1 - \beta^2(t)}\,$.

In Section 2 we set up the notions of simultaneity and proper length for a body in arbitrary motion, that will be the suitable framework to study the spaceships thought experiment that we comment in Section 3, both in its mild and though variants, and in Section 4 we discuss the Redzic's \cite{Redzic2008} paradoxical conclusions. 

\section{Preliminar notions and definitions \label{S2}}
We shall assume that $R$, $F$ and the thread in between lay along the $X$ axis of $\mathcal{S}$, between $x_R=0$ and $x_F=h$ at $t=0$, and we shall ignore the transverse dimensions. Each thread point will be labeled by its initial $X$ coordinate  $0 \leq \xi \leq h\,$ and its worldline, which we assume timelike, is 
\begin{equation}  \label{e1}
x^a = x^a(T,\xi) \,, \qquad \qquad a=0,1
\end{equation}
in $\mathcal{S}$ Lorentzian coordinates. The time parameter $T$ is ticked by some local clock traveling with the point $\xi$. 

As two different points on the thread will never intersect, (\ref{e1}) establishes a one-to-one correspondence 
$$ \,\xi^a= (T,\xi) \longleftrightarrow x^b = (t,x)\,$$ 
between a domain in $\mathbb{R}^2$ and the 1+1 spacetime region spanned by the thread and so $(T,\xi)$ is a system of curvilinear 
(non-inertial) coordinates on that region, which we shall refer to as {\em comoving coordinates}. 
The invariant interval $\,\D s^2 = - \D t^2 + \D x^2\,$ in these coordinates reads
\begin{equation}  \label{e1a}
 \D s^2 = g_{ab}\,\D \xi^a\,\D \xi^b \,, \qquad \quad 
g_{ab}(T,\xi) = -\frac{\partial t}{\partial \xi^a}\frac{\partial t}{\partial \xi^b} + \frac{\partial x}{\partial \xi^a}\frac{\partial x}{\partial \xi^b} 
\end{equation}
(we use natural units so that $c=1$).

\subsection{Proper time \label{S2.1}}
According to the clock hypothesis, a standard clock comoving with the point $\xi$ in the thread ticks the proper time parameter of that worldline. In comoving coordinates the worldline equation is $\,\xi^1= \xi\,$, constant and $\,\xi^0=T\in\mathbb{R}\,$ is the parameter; the velocity vector is $\,v^b=\delta_0^b\,$ and proper time is connected with the invariant interval by
$\, \D\tau^2 = - \D s^2 = g_{ab} v^a v^b\,\D T^2 = - g_{00} \,\D T^2 \,$, hence the rates of local proper time and coordinate time are related by
\begin{equation}  \label{e2}
 \D \tau = \sqrt{- g_{00}(T,\xi)}\,\D T
\end{equation}
and, as $\D\tau^2 >0\,$ (the worldline is timelike), $\,g_{00} \,$ must be negative.
The proper velocity of the point $\xi$ in the thread is thus 
\begin{equation}  \label{e3}
u^a(T,\xi) = \frac{\partial x^a(T,\xi)}{\partial \tau} =\frac1 {\sqrt{- g_{00}}}\,\delta_0^a
\end{equation}

\subsection{Simultaneity (at a distance)\label{S2.2}}
Consider two neighboring points in the thread, $\,\xi\,$ and $\xi + \D \xi\,$, and three close events, $A$, $B$ and $C$, as depicted in the figure:\\[2ex]
\begin{minipage}{11cm}
A radar signal, $(-)$, is emited from $A = (\xi+\D\xi, T +\D T_-)$ that is received at $B = (\xi, T)$ and reflected back as the radar signal   $(+)$ that is finally received at $C = (\xi+\D\xi, T +\D T_+)$.

$t^\dagger$ and $x^\dagger$ are the time and space axes of the inertial system $\mathcal{S}^\dagger_B\,$ wich is instantaneously comoving with the thread point $\xi$ at the event $B$. The $\mathcal{S}^\dagger_B\,$ time axis is the world line of its origin of coordinates, it is the straight line tangent to the worldline $\xi$ at $B$ and it is parallel to $u^a_B$. On its turn, the $\mathcal{S}^\dagger_B\,$ space axis is Minkowski orthogonal to the time axis.
\end{minipage}
\ \hspace*{.5em} \
\begin{minipage}{4.5cm}
\begin{center}
\includegraphics[width=4.5cm]{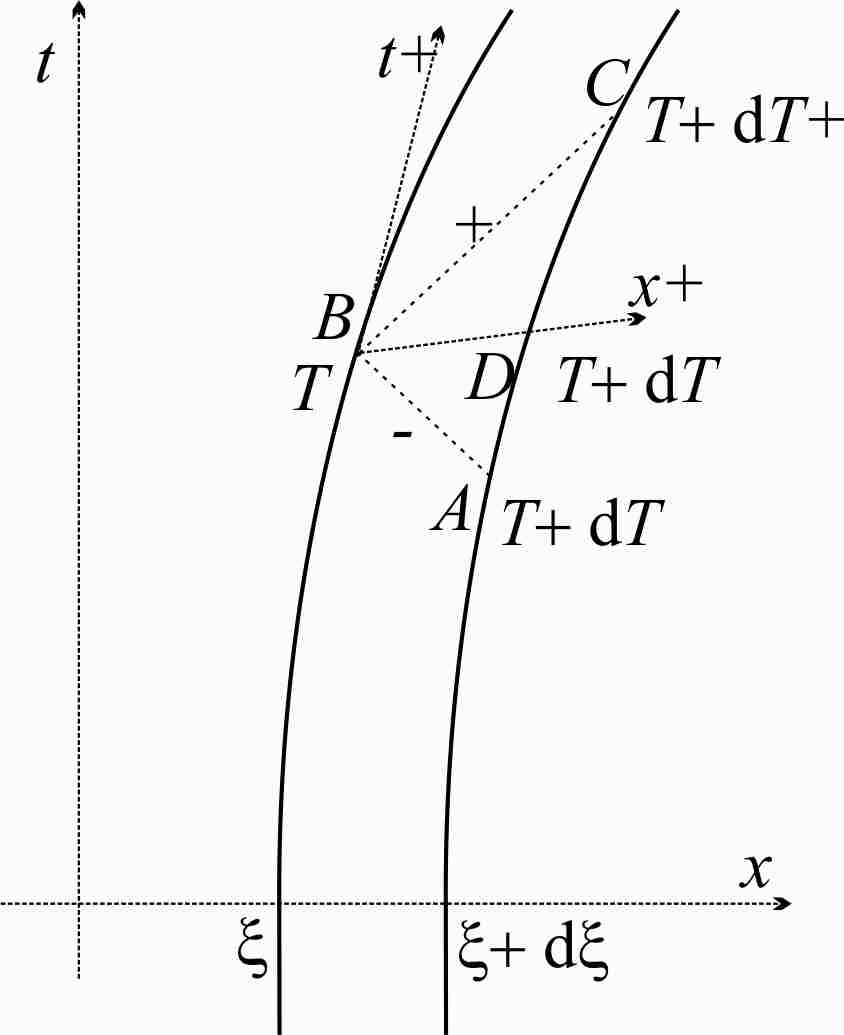}
\end{center}
\end{minipage}

\vspace*{2ex}
As the lines $(+)$ and $(-)$ represent light rays, $\, \D s^2_{AB} = \D s^2_{BC} = 0\,$, and we have 
$$ g_{11}\,\D \xi^2 + 2\,g_{01}\,\D \xi\,\D T_\pm + g_{00}\,\D T_\pm^2 = 0 $$
whence it follows that \cite{Landau}
\begin{equation}  \label{e4}
 \D T_\pm = \frac1{g_{00}}\,\left( - g_{01} \,\D\xi \mp \sqrt{\left[g_{01}^2 - g_{00} g_{11}\right] \,\D\xi^2} \right)
\end{equation}
The whole scheme reminds the telegrapher protocol \cite{Galison} for clock synchronization  proposed by Einstein \cite{Einstein1905}, so that we can define {\em local simultaneity} with respect to the thread as follows:
\begin{quote}
The events $B$ and $D$ in close worldlines, respectively $\xi$ and  $\xi + \D\xi\,$ are simultaneous if, and only if, their time coordinates $\,T\,$ and  $\,T + \D T\,$ satisfy
\begin{equation}  \label{e5}
\D T = \frac12\,\left(\D T_+ + \D T_-\right) = - \frac{g_{01}}{g_{00}}\,\D \xi \,,
\end{equation} 
\end{quote}
that is two neigboring events with coordinates $\xi^a$ and $\xi^a+\D \xi^a$ are locally simultaneous according to the thread's frame whenever
\begin{equation}  \label{e5a}
g_{0 a}\,\D \xi^a = 0 \qquad \mbox{or, equivalenlty,} \qquad u_a(\xi) \,\D\xi^a = 0 
\end{equation}
where we have 
taken  $u_a = g_{ab} u^b$, with the proper velocity given by (\ref{e3}).

It is worth to stress here that the events that are simultaneous to $B$ according to the locally comoving inertial frame $\mathcal{S}^\dagger_B$ satisfy $t^\dagger=0$ or, in general coordinates, $\,\left(x^a - x^{a\,\dagger}_B\right)\,u_{B\, a} = 0\,$, whose infinitesimal expression is nothing but (\ref{e5a}). Hence the {\em local simultaneity} here defined amounts to simultaneity with respect to the local instantaneously comoving inertial frame. 

By invoking transitivity we can extend the simultaneity relation (\ref{e5a}) connecting close events to any pair of events. The latter relation is a total differential equation which, in 1+1 dimensions, is always integrable, i. e. there exist a function $f$ and an integrating factor $\mu$ such that $u_a(\xi) \,\D\xi^a = \mu(T,\xi)\,\D f\,$. 
The solutions are the {\em local simultaneity curves}, each one consisting of a class of events that are locally simultaneous, each event with its neighbors and, by transitivity, to the whole class.

\subsection{Proper length. Born distance \label{S2.3}}
As the radar signal $ABC$ travels from the thread point $\xi + \D\xi$ to $\xi$ and back, it takes a lapse $\Delta T = \D T_+ - \D T_-\,$, which could be a taken as a radar measure of the distance between those thread points. However $\,\Delta T\,$ is a lapse of coordinate time and it depends on the coordinate clocks. For the sake of invariance we translate it into proper time (standard clock measurements) and obtain
$$ \Delta \tau = \sqrt{-g_{00}}\,\Delta T = 2 \,\sqrt{\left(g_{11} - \frac{g_{01}^2}{g_{00}}\right) \,\D\xi^2  } $$
which is the {\em infinitesimal radar distance} between those points (recall that $c=1$)
\begin{equation}  \label{e6}
\D l^{\dagger 2} = \left(g_{11} - \frac{g_{01}^2}{g_{00}}\right) \,\D\xi^2 
\end{equation} 
To compare this with Born's definition of distance\footnote{Born's definition is for 3+1 dimensions and here we refer to its specialization to 1+1 spacetimes} \cite{Born1909}, \cite{Bel1994} between points moving along a flow with unit proper velocity field $u^a\,$, namely
\begin{equation}  \label{e6a}
 \D l_B^2 = \hat{g}_{ab}\,\D \xi^a\,\D\xi^b \,, \qquad {\rm with} \qquad \hat{g}_{ab} =  g_{ab} + u_a u_b \,, 
\end{equation}
it suffices to substitute $\,\displaystyle{u_a = \frac{g_{0a}}{\sqrt{-g_{00}}} } \,$ in the above expression to obtain that $\,\D l^\dagger = \D l_B\,$.

An important feature to bear in mind is that the length (\ref{e6}) between the thread points $\xi$ and $\xi+\D\xi\,$ might be time dependent: the outcome of two measures of the same segment at different times are not necessarily equal.

We can also see the connection of Born's distance (\ref{e6}) with the usual definition of proper length measured in a rest frame. Consider the event $D$ in the figure above that, as seen before, is $\mathcal{S}^\dagger_B$-simultaneous with $B$. The proper length of the thread segment $BD$ is the outcome of the measurement in the inertial frame $\mathcal{S}^\dagger_B$ comoving with $BD$, i. e. $\,|x_D^\dagger - x_B^\dagger|\,$ and, as $\,t_D^\dagger=t_B^\dagger\,$, we have that 
$$ |x_D^\dagger - x_B^\dagger|^2 = \D s_{BD}^2 $$
In comoving coordinates the invariant interval is 
$$  \D s_{BD}^2 = g_{00}\,\D T^2 +  2\,g_{01}\,\D T \,\D\xi  +  g_{11}\,\D \xi^2  $$
and, including (\ref{e5}) that relates $\D T$ and $\D\xi$ for locally simultaneous events, we arrive at $\,\D l^{\dagger 2} = \D s_{BD}^2 \,$, which proves that {\em Born's distance} and {\em local instantaneous proper length} are the same.

To extend this definition, which is valid for an infinitesimal segment of the thread, to a finite piece of it, we decompose the thread in infinitesimal segments and then add their infinitessimal proper lengths but, as the length of each segment might depend on time, we shall be sure that all these infinitesimal lengths correspond to the ``same time'', i. e. to simultaneous configurations of the segments we are adding.
Thus we can define the proper length of the piece of thread included between $0 \leq \xi \leq \Xi$ as 
\begin{equation}  \label{e7}
L^\dagger = \int_\mathcal{C} \D l^\dagger
\end{equation}
where $\mathcal{C}$ is a simultaneity curve crossing the worldlines from $\xi=0$ to $\Xi\,$. 
For obvious reasons this length may be variable because both simultaneity curves and Born's length depend on the local motion of the thread points. 

This suggests a definition of {\em inextensibility}; a thread is inextensible iff it only admits motions that preserve Born's distance, i. e. its Lie derivative along the velocity vector vanishes: $\quad\mathcal{L}_{\mathbf{u}} \hat{g}_{ab}=0\,$.

The above definition of proper length has the expected feature of being additive, that is if we consider the thread as made of two pieces, the total length (under any motion) is the sum of the lengths of each part.

It is worth noticing that defining the proper length of an object in arbitray motion as the addition of infinitesimal length measurements done in locally comoving inertial frames is an instance of the Hypothesis of Locality \cite{Mashhoon} mentioned at the Introduction.

\section{The Dewan-Beran thought experiment \label{S3}}
We now apply the notions and definitions advanced so far to the thread between two spaceships in Dewan and Beran thought experiment and, following Redzic \cite{Redzic2008}, we distinguish a tough variant, in which the spaceships keep accelerating forever, and a mild variant, when the ships run out of gas and eventually reach a state of uniform motion. 

As the definitions of proper length and simultaneity proposed in Section \ref{S2} depend on the state of local motion of the thread, for the sake of simplicity we shall assume that each point in the thread has the same  uniformly accelerated motion with respect to $\mathcal{S}$ as both ends, $R$ and $F$. This is a simplified and rather unrealistic version of the ``true motion'' since it presumes that the push by $R$ and the pull by $F$ are instantaneously transmitted (according to $\mathcal{S}$) to all thread points, which implies a signal propagating faster than light.  
A more realistic and consistent picture should include an elastic model of the thread, but this is beyond the scope of the present study and  will be the object of future work.

\subsection{The tough variant \label{S3.1}}
The rear and front ships,  and the thread points as well, move linearly at a constant proper acceleration starting at $t=0$ and the worldlines in $\mathcal{S}$-Lorentzian coordinates are
\begin{equation}  \label{e8}
x = \xi + \frac1a\,\left[\cosh(a\tau) - 1\right] \,, \qquad \qquad t =   \frac1a\, \sinh(a\tau)
\end{equation}
where $\,0 \leq \xi \leq h\,$ is the starting abscissa in $\mathcal{S}$, $\xi=0$ and $\xi = h$  respectively correspond to the $R$ and $F$ ships, and $\tau$ is proper time on each world line. 

In these adapted coordinates the invariant interval (\ref{e1a}) is 
\begin{equation}  \label{e9}
\D s^2 = - \D \tau^2 + 2\,\sinh(a\tau)\,\D\tau\,\D\xi + \D\xi^2 
\end{equation}
The presence of the cross term $\,\D\tau\,\D\xi \,$ indicates that equal $\tau\,$ does not mean simultaneity and, according to (\ref{e5a}) the differential equation for local simultaneity curves is
\begin{equation}  \label{e10}
 - \D\tau + \sinh(a\tau)\,\D\xi = 0 
\end{equation}
which can be easily solved to obtain
\begin{equation}  \label{e11}
 \xi = \xi_0 + \frac1a\,\ln\frac{e^{a\tau}-1}{e^{a\tau}+1} \,, \qquad \qquad \qquad \tau > 0
\end{equation}
where $\xi_0$ is an integration constant. Inverting it we obtain the explicit equation for simultaneity curves in a $(\xi,\tau)$ diagram
\begin{equation}  \label{e11a}
 \tau(\xi) =  \frac1a\,\ln\frac{e^{a\xi_0}+e^{a\xi}}{e^{a\xi_0}-e^{a\xi}} \,, \qquad \qquad \qquad \tau > 0
\end{equation}

Notice that (\ref{e11}) implies $\,\xi < \xi_0\,$, which means that the domain of $\tau(\xi)$ is bounded from above and that each simultaneity curve has a horizon. 

If we are interested in the locus of events that are locally simultaneous with 
the event $\,(\tau_1,0)\,$ on the ship $R$, the condition $\,\tau_1=\tau(0)\,$ and (\ref{e11a}) allow to determine the integration constant 
\begin{equation}  \label{e12}
  \xi_0 =  - \frac1a\,\ln\frac{e^{a\tau_1}-1}{e^{a\tau_1}+1} \,, \qquad {\rm or} \qquad 
	e^{-a\xi_0} = \tanh\left(\frac{a\tau_1}{2}\right)
\end{equation}
whence it follows that $\xi_0$ depends on $\tau_1$, the time on the rear ship, and $\xi_0(\tau_1) > 0\,$ whenever $\tau_1 > 0\,$. Substituting it in equation (\ref{e11a}) we have
\begin{equation}  \label{e12a}
 \tau(\xi,\tau_1) =  \frac1a\,\ln\frac{1 + e^{a\xi} \tanh(a\tau_1/2)}{1 - e^{a\xi} \tanh(a\tau_1/2)} \,, 
\end{equation}
which is the proper time on the place $\xi$ of the event that is simultaneous with $(\tau_1,0)\,$. In other words, the events $\, ( \tau(\xi,\tau_1),\xi) \,$ and $\,(\tau_1,0)\,$ are locally simultaneous.

It also follows from (\ref{e11}) and (\ref{e11a}) that, if the local simultaneity curve starts too late at the rear end of the thread, then it might be that the front end is beyond the horizon. This happens for $\tau_1>\tilde\tau_1$, the solution of $\tau(h,\tilde\tau_1) = \infty\,$ which, including (\ref{e12a}), yields
\begin{equation}  \label{e13}
  1 - e^{a h} \,\tanh\left(\frac{ a\tilde\tau_1}{2}\right) = 0 \qquad {\rm or} \qquad 
	\tilde\tau_1 = \frac1a\,\ln \frac{1+e^{-a h}}{1-e^{-a h}}
\end{equation}

Furthermore, if $(\tau_2,h)\,$ is the event on $F$ that is locally simultaneous with the event $(\tau_1,0)\,$ on $R$, we easily obtain from (\ref{e12a}) that
\begin{equation}  \label{e13a}
  \tau_2 =  \frac1a\,\ln\frac{1 + e^{a h} \tanh(a\tau_1/2)}{1 - e^{a h} \tanh(a\tau_1/2)}  \,,\qquad {\rm or} \qquad  \tanh\left(\frac{ a \tau_2}{2}\right) = e^{a h} \,\tanh\left(\frac{ a \tau_1}{2}\right) 	
\end{equation}	
It is obvious that $\tau_2$ goes to infinity when $\tau_1$ approaches $\tilde\tau_1\,$, i. e. $\xi_0 $ approaches $h$, and simultaneity curves that start at the rear end later than $\tilde\tau_1$ never reach the thread's front end.

\begin{figure}[ht]
\begin{center}
{\includegraphics[width=15cm]{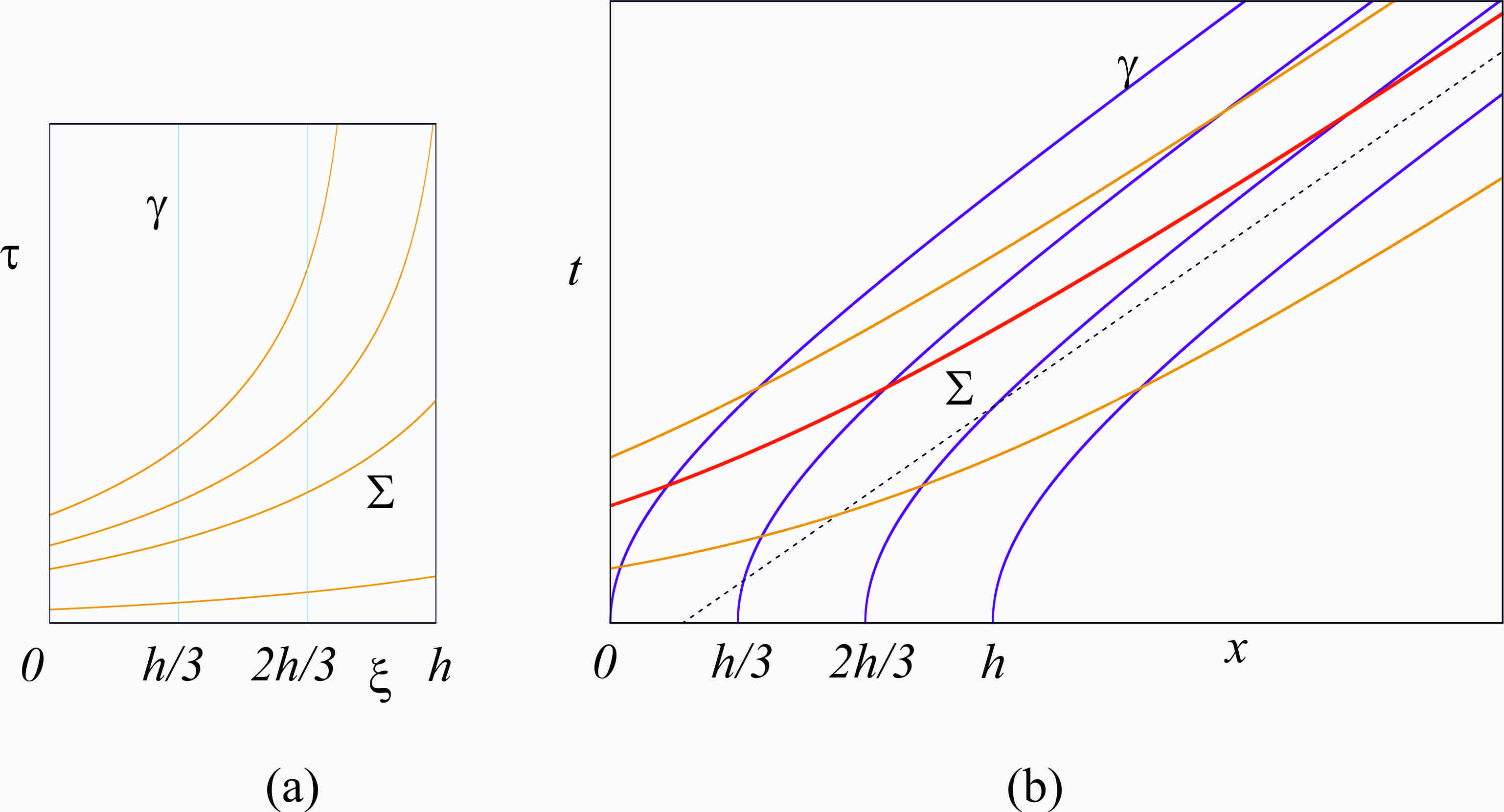}}
\end{center}
\caption{Tough variant: Worldlines ($\gamma$) of some thread points and some simultaneity lines ($\Sigma$): (a) in adapted coordinates and  (b) in Lorentzian $\mathcal{S}$ coordinates }
\end{figure}

As for the  proper length of an infinitesimal piece of thread, from (\ref{e6}) and (\ref{e9}) it follows that
\begin{equation}  \label{e14}
 \D l^\dagger = \cosh(a\tau)\,\D\xi
\end{equation}
Since the total length is defined as the integral (\ref{e7}) over a local simultaneity curve for $0 \leq \xi \leq h$, it only makes sense for $\tau_1 < \tilde\tau_1\,$ and, including (\ref{e12a}), we have that
$$l^\dagger =  \int_0^h \frac{1 + e^{2 a\xi}\,\tanh^2(a\tau_1/2)}{1 -  e^{2 a\xi}\,\tanh^2(a\tau_1/2)}  \,\D \xi $$
For later rear times, $\tau_1 > \tilde\tau_1\,$, the simultaneity curve never intersects the front's world line and therefore the definition advanced here cannot apply.

After a short calculation the latter integral yields
$$ l^\dagger  = h + \frac1a\, \ln \frac{1 - \tanh^2(a\tau_1/2)}{1 - \tanh^2(a\tau_2/2)}   $$
that, including (\ref{e13a}) can be written in terms of the rear time $\tau_1$,
\begin{equation}  \label{e14a}
 l^\dagger  = h + \frac1a\, \ln \frac{1 - \tanh^2(a\tau_1/2)}{1 - e^{2 a h} \tanh^2(a\tau_1/2)} 
\end{equation}
which goes to infinity when $\tau_1$ approaches $\tilde\tau_1\,$.

We could likewise put $l^\dagger\,$ as a function of the front time $\tau_2$ and write
\begin{equation}  \label{e14b}
 l^\dagger  = h + \frac1a\, \ln \frac{1 - e^{-2 a h}\,\tanh^2(a\tau_2/2)}{1 - \tanh^2(a\tau_2/2)} 
\end{equation}
which is unbounded but finite for all $\tau_2\geq 0\,$.

Further simplification yields an expression that will be useful later 
\begin{equation}  \label{e14aa}
 l^\dagger  = h + \frac2a\, \ln \frac{\cosh(a\tau_2/2)}{\cosh(a\tau_1/2)}   
\end{equation}
where $\tau_2$ and $\tau_1$ are related by (\ref{e13a}). 

\subsection{The mild variant \label{S3.2}}
Assume that the spaceships are identical and that both run out of gas after a proper time $\tau^\ast$. Assume also that the thread points move in the same manner as the front and rear ships. Their individual worldlines are given piecewise by:
\begin{equation}  \label{e18}
\left. \begin{array}{ll}
       \displaystyle{x = \xi + \frac1a\,\left[\cosh(a\tau) - 1\right] \,, \qquad \quad t =   \frac1a\, \sinh(a\tau)} & 0 \leq \tau \leq \tau^\ast \\
			 \displaystyle{x = x^\ast + (\tau-\tau^\ast)\,\sinh(a\tau^\ast)  \,, \qquad  t =  t^\ast + (\tau-\tau^\ast)\,\cosh(a\tau^\ast) }  & \tau^\ast < \tau 
			\end{array}  \right\}
\end{equation}
where $\, x^\ast = x(\tau^\ast) \,$ and $\,  t^\ast =  t(\tau^\ast)  \,$.
The worldlines have continuous derivatives up to the first order.

The invariant interval (\ref{e1a}) is, in adapted coordinates, 
\begin{equation}  \label{e19}
\D s^2 = - \D \tau^2 + 2\,S(\tau)\,\D\tau\,\D\xi + \D\xi^2  
\end{equation}
with $ \quad S(\tau) = \sinh(a\tau)\,, \quad {\rm if} \quad 0 \leq \tau \leq \tau^\ast \,,\qquad {\rm or} \qquad 
S(\tau) = \sinh(a\tau^\ast) \,, \quad {\rm if} \quad    \tau^\ast  < \tau \,$.

Local simultaneity curves satisfy $\,\D\tau = S(\tau)\,\D\xi \,$, whose general solutions is
\begin{equation}  \label{e20}
\xi(\tau,\tau_1) = 
\left\{ \begin{array}{ll}
       \displaystyle{ \xi_0(\tau_1) + \frac1a\,\ln\frac{e^{a\tau} -1}{e^{a\tau} +1} }\,, & \qquad 0 \leq \tau \leq \tau^\ast \\[1ex]
			 \displaystyle{  \xi^\ast(\tau_1) + \frac{\tau-\tau^\ast}{\sinh(a\tau^\ast)}  \,, }  & \qquad \tau^\ast < \tau 
			\end{array}  \right.
\end{equation}
where $\, \displaystyle{\xi^\ast = \xi(\tau^\ast,\tau_1) = \xi_0(\tau_1) + \frac1a\,\ln\frac{e^{a\tau^\ast} -1}{e^{a\tau^\ast} +1}  }\,$ or, its inverse
\begin{equation}  \label{e21}
\tau(\xi,\tau_1) =
\left\{ \begin{array}{ll}
       \displaystyle{   \frac1a\,\ln\frac{1 + e^{a\xi} \tanh(a\tau_1/2)}{1 - e^{a\xi} \tanh(a\tau_1/2)} \,,}\,, & \qquad 0 \leq \xi \leq \xi^\ast(\tau_1) \\[1ex]
			 \displaystyle{ \tau^\ast + \left[\xi-\xi^\ast(\tau_1)\right]\,\sinh(a\tau^\ast) \,,}  & \qquad \xi^\ast(\tau_1) < \xi 
			\end{array}  \right.
\end{equation}
where equation (\ref{e12}) has been included. Since $\,\displaystyle{ \frac{\D \xi}{\D \tau} > 0\,}$ both, $\xi(\tau,\tau_1)$ and its inverse $\tau(\xi,\tau_1)\,$, are increasing functions of $\tau\,$.

\begin{figure}[ht]
\begin{center}
{\includegraphics[width=15cm]{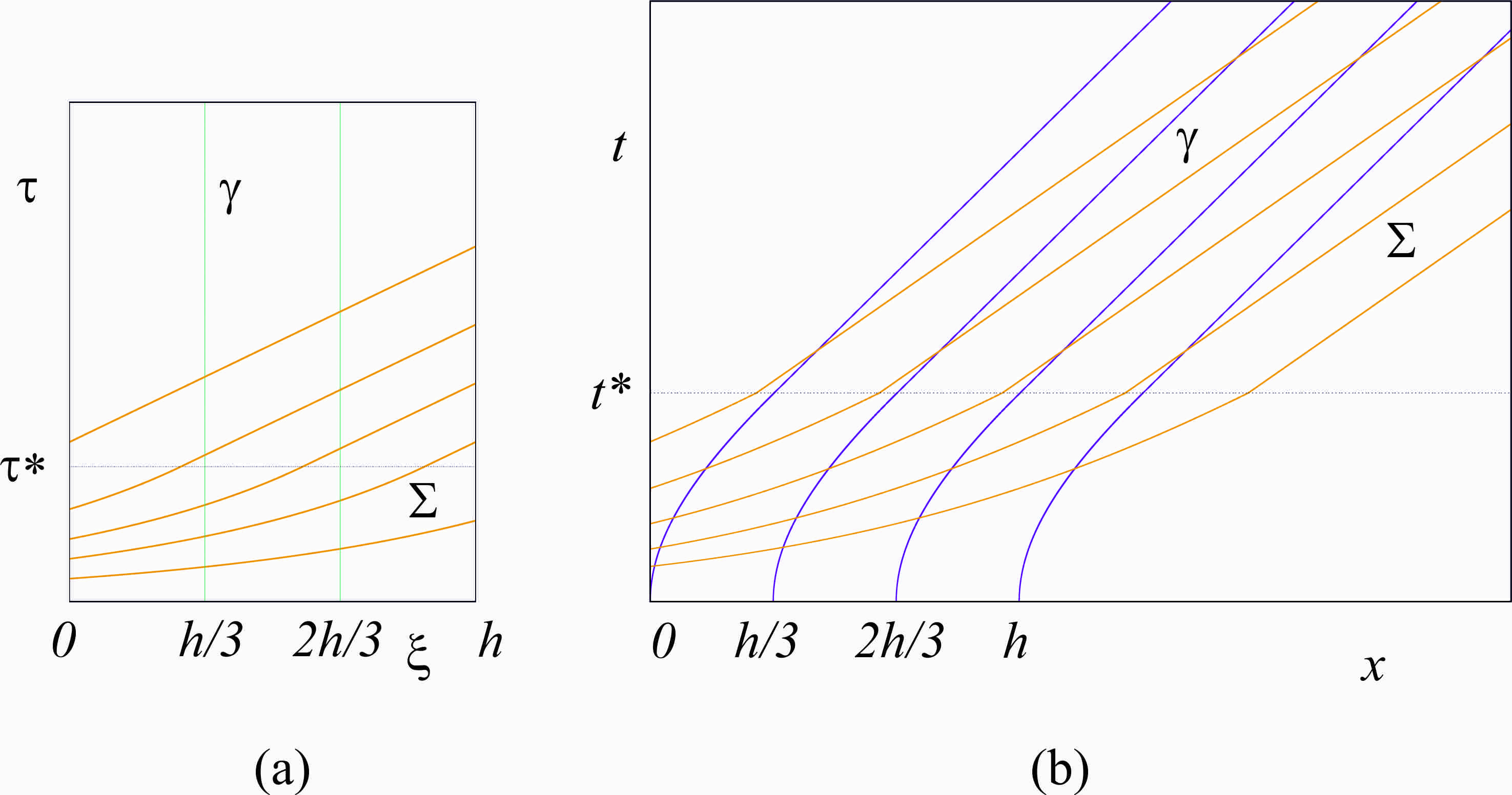}}
\end{center}
\caption{Mild variant: Worldlines ($\gamma$) of some thread points and some simultaneity lines ($\Sigma$): (a) in adapted coordinates and  (b) in Lorentzian $\mathcal{S}$ coordinates }
\end{figure}

Notice that $\xi^\ast(\tau_1)$ lays always before the horizon, i. e. $\xi^\ast (\tau_1)< \xi_0(\tau_1)\,$. Moreover, if $\,0 < \xi^\ast(\tau_1) \leq h\,$, the local simultaneity curve is a matching of the two branches (\ref{e21}), otherwise it consists of only one branch. 

From (\ref{e19}) we easily obtain that the infinitesimal Born's length (\ref{e6}) is $\,\D l^\dagger = \gamma(\tau)\,\D\xi \, $, with
\begin{equation}  \label{e22}
\gamma(\tau) = \cosh(a\tau) \quad {\rm if} \quad 0 \leq \tau \leq \tau^\ast \qquad {\rm or} \qquad 
\gamma(\tau) = \cosh(a\tau^\ast) \quad {\rm if} \quad    \tau^\ast  < \tau 
\end{equation}
As for the total proper length of the thread, we have to calculate the integral (\ref{e7}) over local simultaneity curves, and we distinguish three cases: (a) $\,\xi^\ast(\tau_1) \leq 0 \quad $,   (b) $\,0 \leq \xi^\ast(\tau_1) \leq h\quad$, and  (c) $\,h \leq \xi^\ast(\tau_1)\,$.

Including (\ref{e12}), the above expression for $\xi^\ast\,$ implies that
\begin{equation}  \label{e22a}
 \tanh\left(\frac{a\tau^\ast}{2}\right)= e^{a \xi^\ast}\,\tanh\left(\frac{ a \tau_1}{2}\right) 
\end{equation}
As a consequence, if we denote by $\bar\tau_1$ the rear time which is locally simultaneous with the front time $\tau^\ast$, we have that
$$ \tanh\left(\frac{a\tau^\ast}{2}\right)= e^{a h}\,\tanh\left(\frac{ a \bar\tau_1}{2}\right)  $$

Thus the proper length in each case is:
\begin{description}
\item[\underline{$ \xi^\ast(\tau_1) < 0$}] or $\tau^\ast \leq \tau_1$, which implies that $\tau(\xi,\tau_1) > \tau_1\,$ because it is an increasing function. Hence $\,\gamma(\tau) = \cosh(a\tau^\ast) \,$ is constant and $\, l^\dagger = h \,\cosh(a\tau^\ast) \,$, 
i. e. the length measured by the inertial system $\mathcal{S}^\ast\,$ comoving with the whole thread from $\tau^\ast$ on, or the proper length in the usual special relativity sense.
\begin{equation}  \label{e24a}
 l^\dagger(\tau_1) = h \,\cosh(a\tau^\ast)   \,, \qquad \tau^\ast \leq \tau_1
\end{equation}
\item[\underline{$ 0 \leq \xi^\ast(\tau_1) \leq h$}] or $\bar\tau_1  \leq  \tau_1 \leq \tau^\ast $, hence
 the simultaneity curve starting at $\,(\tau_1,0)\,$ is contributed by the two sections in (\ref{e20}) and the total proper length splits as 
$$ l^\dagger = \int_0^{\xi^\ast} \cosh(a\tau) \,\D\xi + \int_{\xi^\ast}^h \cosh(a\tau^\ast) \,\D\xi $$
The second term is $\, (h - \xi^\ast) \cosh(a\tau^\ast)\,$, i. e. the ordinary inertial proper length for the thread from $\xi^\ast$ to $h$, whereas the first term amounts to the tough variant result (\ref{e14aa}) for a thread spanning from 0 till $\xi^\ast$.   
The total proper length is thus
\begin{equation}  \label{e24c}
 l^\dagger (\tau_1) = h \,\cosh(a\tau^\ast)  + \frac2a\, \ln \frac{\cosh(a\tau^\ast/2)}{\cosh(a\tau_1/2)} +  \xi^\ast\,\left[ 1 -\cosh(a\tau^\ast) \right]   
\end{equation} 
\item[\underline{$h \leq \xi^\ast(\tau_1)$}]  or $\tau_1 \leq \bar\tau_1 \leq \tau^\ast$
The section of simultaneity curve between $\, 0 \leq \xi \leq h \,$ is the same as in the tough variant and the outcome for the total proper length is the same as well, i. e. equation (\ref{e14aa}).
\begin{equation}  \label{e24b}
 l^\dagger(\tau_1)   = h  + \frac2a\, \ln \frac{\cosh(a\tau_2/2)}{\cosh(a\tau_1/2)} 
\end{equation}
with $\tau_2$ given by (\ref{e13a}).
\end{description}

It can be easily checked that the derivative of $\,l^\dagger(\tau_1)\,$  is positive, hence the proper length increases from $\,l^\dagger(0) = h\,$ to reach the constant value $\, h \,\cosh(a\tau^\ast) \,$ for rear times later that $\tau^\ast\,$. Furthermore, by taking the suitable lateral limits, we find that the three stretches (\ref{e24a}-\ref{e24b}) match smoothly with each other and the proper length $\,l^\dagger(\tau_1)\,$ has continuous first order derivatives. 

\section{A comment on Redzic's paradox  \label{S4}}
Our conclusion is that the thread is always longer than $h$, its proper length increases up to $h\,\cosh(a\tau^\ast)\,$, in the mild variant, or boundlessly in the tough variant; therefore the thread breaks apart fatally in the tough variant and in the mild variant as well, provided that the Lorentz factor $\,\cosh(a\tau^\ast)\,$ is large enough. This outcome is less shocking than Redzic's  conclusion that ``\ldots the final outcome need not be fatal in the tough variant of the riddle'' \cite{Redzic2008}. 

Redzic defines the instantaneous length of the moving thread as the distance between two events $E_1$ and $E_2$, respectively on the rear and front ends of the thread, that are simultaneous according to an inertial frame $\mathcal{S}^\prime$ that is instantaneously comoving with the front ship at the event $E_2$. In the mild variant, when the engines stop at $\tau^\ast$, he obtains the same final length as Dewan and Beran \cite{Dewan1959}, that is $h$ times the Lorentz factor $\cosh(a\tau^\ast)\,$, but in the tough variant of the experiment the threads length, according to his definition, does not increase boundlessly. 

In our notation the inertial $\mathcal{S}$ coordinates of these events is
\begin{eqnarray*}
\underline{E_2} & \quad & x(h,\tau_2) = h + \frac1a\,\left[\cosh(a\tau_2) - 1 \right] \,, \qquad 
t(h,\tau_2) = \frac1a\,\sinh(a\tau_2) \\[2ex]
\underline{E_1} & \quad & x(0,\tau_1) =  \frac1a\,\left[\cosh(a\tau_1) - 1 \right] \,, \qquad \qquad
t(h,\tau_1) = \frac1a\,\sinh(a\tau_1)
\end{eqnarray*}
The time axis of the comoving inertial frame $\mathcal{S}^\prime$, i. e. the proper velocity at the event $E_2$ is
$$ u^x(h,\tau_2) = \sinh(a\tau_2) \,,\qquad  u^t(h,\tau_2) = \cosh(a\tau_2)  \,, $$
and the condition of  $\mathcal{S}^\prime$-simultaneity of $E_1$ and $E_2$, $\quad \left[ x^\mu(h,\tau_2) - x^\mu(0,\tau_1)\right]\,u_\mu(h,\tau_2) = 0 \,$ yields
\begin{equation}  \label{e25}
 \sinh a(\tau_2-\tau_1) = h \,a\,\sinh (a\tau_2)  
\end{equation}
The vector $r^\mu$, with $r^x = \cosh(a\tau_2)\,$ and $\,r^t = \sinh(a\tau_2)\,$, completes an orthonormal spacetime base and the distance $L$ between $E_1$ and $E_2$ can be obtained either as $\,L =\left|r_\mu \,\left[ x^\mu(h,\tau_2) -  x^\mu(0,\tau_1)\right] \right|\,$ or as the square root of the invariant interval $E_1E_2$. After a short calculation that includes (\ref{e25}) we obtain that
$$ L = \frac1a + h\,\left[\cosh(a\tau_2) - \sqrt{\cosh^2(a\tau_2) +\frac1{h^2a^2} - 1 } \right] $$
which has the finite limit $1/a$ when $\tau_2\rightarrow \infty$.

The reason of Redzic's paradoxical result is that he takes the instantaneous proper length of the moving thread as the distance between two events $E_1$ and $E_2$, respectively on the rear and front ends of the thread, that are simultaneous according to an inertial frame $\mathcal{S}^\prime$ which is instantaneously comoving with the front ship at event $E_2$. In the mild variant, for $\tau_2$ large enough, both ends of thread move at the same speed (with respect to $\mathcal{S}$) and are at rest as seen from $\mathcal{S}^\prime\,$, and Redzic's proper length coincides with the usual special relativistic definition. However, in the tough variant of the experiment it is obvious that the thread is not instantaneously at rest in $\mathcal{S}^\prime$ because, according with $\mathcal{S}$-time $E_1$ is prior to $E_2$, and the speed of the rear end at $E_1$ is less than the front speed at $E_2$.

However this definition is somewhat asymmetric in that the front end is preferred to the rear end. Instead of taking the thread's length as the proper distance between two events, $E_1$ and $E_2$ one at each end, that are simultaneous with respect to the inertial frame $\mathcal{S}^\prime$ which sees the front at rest at event $E_2$, with equal legitimacy one could choose the simultaneity with respect to the inertial frame $\mathcal{S}^\prime_1$ which sees the rear at rest at event $E_1$. Then the relation connecting the proper time coordinates of $E_1$ and $E_2$ would be
$$ h a\,\sinh(a\tau_1) = \sinh a(\tau_2-\tau_1)   $$
and the distance between $E_1$ and $E_2$ as measured by $\mathcal{S}^\prime_1$ is
$$ L_1 = h\,\cosh (a\tau_1) -\frac1a + \sqrt{h^2 \,\sinh^2 (a\tau_1) + \frac1{a^2} }  $$
which is unbounded for $\tau_1\rightarrow \infty$\,.

Furthermore Redzic's definition has not the property of additivity that one would expect for a length.

\section{Conclusion}
We have drawn upon the Dewan-Beran thought experiment, also known as Bell's spaceships paradox, to illustrate the difficulties in extending the notion of proper length of a body (a thread) in arbitrary motion. As a rule there is no inertial reference frame instantaneously comoving with the body, i. e. such that the whole body is instantaneously at rest with respect to it, and we must resort to the so called Hypothesis of Locality \cite{Mashhoon}. The definition of proper distance we adopt here is the sum of the Born lengths of the infinitesimal parts of the body, simultaneously considered, according with the notion of local simultaneity \cite{Born1909},\cite{Landau}.

When applying these ideas to the thread that connects the spaceships $R$ and $F$ in the Dewan-Beran thought experiment, we have assumed that all thread points have the same uniformly accelerated motion as the ships $R$ and $F$. We admit that this is an oversimplification and that  assuming that the pull of $F$ and the push of $R$ are elastically transmitted to the other parts in the thread would be more realistic. However, such a treatment would require introducing a relativistic theory of elasticity, which lies beyond the scope of the present study and will be the object of further work. 

\section*{Acknowledgment}
Funding for this work was partly provided by the Spanish MINECO and ERDF (project ref. FPA2016-77689-C2-2-R).

\end{document}